\documentclass[useAMS,usenatbib]{mn2e}

\usepackage{graphicx}
\usepackage{subcaption}
\usepackage{amsmath}
\usepackage{amssymb}
\usepackage{dcolumn}
\usepackage{float}

\usepackage{longtable}
\usepackage{color}
\usepackage{soul}

\definecolor{jaune}{rgb}{1.0, 1.0, 0.0}
\definecolor{vero}{rgb}{1.0, 0.50, 0.75}
\definecolor{alexdu}{rgb}{0.48, 0.40, 0.93}
\definecolor{christi}{rgb}{0.37, 0.62, 0.62}
\definecolor{alexf}{rgb}{0.5, 0.25, 0.25}
\definecolor{new}{rgb}{1.0, 0.0, 0.0}
\definecolor{anew}{rgb}{0.0, 0.0, 1.0}
\definecolor{darkblue}{rgb}{0.0, 0.0, 0.60}

\usepackage[normalem]{ulem}
\newcommand{\stv}{\bgroup\markoverwith{\textcolor{vero}{\rule[0.5ex]{2pt}{0.4pt}}}\ULon}

\def\gtrsim{\mathrel{\hbox{\rlap{\hbox{\lower4pt\hbox{$\sim$}}}\hbox{$>$}}}}
\def\ltsim{\mathrel{\hbox{\rlap{\hbox{\lower4pt\hbox{$\sim$}}}\hbox{$<$}}}}

\usepackage[draft]{hyperref}
\hypersetup{bookmarks=true, colorlinks=true, linkcolor=darkblue, citecolor=darkblue, filecolor=darkblue, urlcolor=darkblue}

\clearpage

\title[UV spectroscopy of NGC 1624-2]{Extreme resonance line profile variations in the ultraviolet spectra of NGC 1624-2: probing the giant magnetosphere of the most strongly magnetized known O-type star
}

\author[A. David-Uraz et al.]{A. David-Uraz,$^{1}$\thanks{E-mail: adu@udel.edu} C. Erba,$^{1}$ V. Petit,$^{1}$ A. W. Fullerton,$^{2}$ F. Martins,$^{3}$\newauthor N. R. Walborn,$^{2}$\thanks{Deceased 22 February 2018.} R. MacInnis,$^{1}$ R. H. Barb\'{a},$^{4}$ D. H. Cohen,$^{5}$ J. Ma{\'\i}z Apell{\'a}niz,$^{6}$\newauthor Y. Naz\'{e},$^{7}$\thanks{F.R.S.-FNRS Research Associate} S. P. Owocki,$^{1}$ J. O. Sundqvist,$^{8}$ A. ud-Doula,$^{9}$ and G. A. Wade$^{10}$
\\
$^{1}$ Department of Physics and Astronomy, University of Delaware, Newark, DE 19716, USA\\
$^{2}$ Space Telescope Science Institute, 3700 San Martin Drive, Baltimore, MD 21218, USA \\
$^{3}$ LUPM-UMR 5299, CNRS \& Universit\'{e} Montpellier, Place Eugene Bataillon, 34095 Montpellier Cedex 05, France \\
$^{4}$ Departamento de F\'{i}sica y Astronom\'{i}a, Universidad de La Serena, Av. Cisternas 1200 Norte, La Serena, Chile\\
$^{5}$ Department of Physics and Astronomy, Swarthmore College, Swarthmore, PA 19081, USA\\
$^{6}$ Centro de Astrobiolog{\'\i}a, CSIC-INTA. Campus ESAC. Camino bajo del castillo s/n. E-28\,692 Villanueva de la Ca\~nada. Spain \\
$^{7}$ Universit\'e de Li\`ege, Quartier Agora (B5c, Institut d'Astrophysique et de G\'eophysique), All\'ee du 6 Ao\^ut 19c, B-4000 Sart Tilman,\\ Li\`ege, Belgium\\
$^{8}$ Instituut voor Sterrenkunde, KU Leuven, Celestijnenlaan 200D, B-3001 Leuven, Belgium\\
$^{9}$ Penn State Scranton, 120 Ridge View Drive, Dunmore, PA 18512, USA\\
$^{10}$ Department of Physics and Space Science, Royal Military College of Canada, PO Box 17000, Stn Forces, Kingston,\\ Ontario K7K 7B4, Canada
}

\begin{document}

%
%
%


\def\jnl@style{\it}
\def\aaref@jnl#1{{\jnl@style#1}}

\def\aaref@jnl#1{{\jnl@style#1}}

\def\aj{\aaref@jnl{AJ}}                   
\def\araa{\aaref@jnl{ARA\&A}}             
\def\apj{\aaref@jnl{ApJ}}                 
\def\apjl{\aaref@jnl{ApJ}}                
\def\apjs{\aaref@jnl{ApJS}}               
\def\ao{\aaref@jnl{Appl.~Opt.}}           
\def\apss{\aaref@jnl{Ap\&SS}}             
\def\aap{\aaref@jnl{A\&A}}                
\def\aapr{\aaref@jnl{A\&A~Rev.}}          
\def\aaps{\aaref@jnl{A\&AS}}              
\def\azh{\aaref@jnl{AZh}}                 
\def\baas{\aaref@jnl{BAAS}}               
\def\jrasc{\aaref@jnl{JRASC}}             
\def\memras{\aaref@jnl{MmRAS}}            
\def\mnras{\aaref@jnl{MNRAS}}             
\def\pra{\aaref@jnl{Phys.~Rev.~A}}        
\def\prb{\aaref@jnl{Phys.~Rev.~B}}        
\def\prc{\aaref@jnl{Phys.~Rev.~C}}        
\def\prd{\aaref@jnl{Phys.~Rev.~D}}        
\def\pre{\aaref@jnl{Phys.~Rev.~E}}        
\def\prl{\aaref@jnl{Phys.~Rev.~Lett.}}    
\def\pasp{\aaref@jnl{PASP}}               
\def\pasj{\aaref@jnl{PASJ}}               
\def\qjras{\aaref@jnl{QJRAS}}             
\def\skytel{\aaref@jnl{S\&T}}             
\def\solphys{\aaref@jnl{Sol.~Phys.}}      
\def\sovast{\aaref@jnl{Soviet~Ast.}}      
\def\ssr{\aaref@jnl{Space~Sci.~Rev.}}     
\def\zap{\aaref@jnl{ZAp}}                 
\def\nat{\aaref@jnl{Nature}}              
\def\iaucirc{\aaref@jnl{IAU~Circ.}}       
\def\aplett{\aaref@jnl{Astrophys.~Lett.}} 
\def\apspr{\aaref@jnl{Astrophys.~Space~Phys.~Res.}}
\def\bain{\aaref@jnl{Bull.~Astron.~Inst.~Netherlands}} 
\def\fcp{\aaref@jnl{Fund.~Cosmic~Phys.}}  
\def\gca{\aaref@jnl{Geochim.~Cosmochim.~Acta}}   
\def\grl{\aaref@jnl{Geophys.~Res.~Lett.}} 
\def\jcp{\aaref@jnl{J.~Chem.~Phys.}}      
\def\jgr{\aaref@jnl{J.~Geophys.~Res.}}    
\def\jqsrt{\aaref@jnl{J.~Quant.~Spec.~Radiat.~Transf.}}
\def\memsai{\aaref@jnl{Mem.~Soc.~Astron.~Italiana}}
\def\nphysa{\aaref@jnl{Nucl.~Phys.~A}}   
\def\physrep{\aaref@jnl{Phys.~Rep.}}   
\def\physscr{\aaref@jnl{Phys.~Scr}}   
\def\planss{\aaref@jnl{Planet.~Space~Sci.}}   
\def\procspie{\aaref@jnl{Proc.~SPIE}}   

\let\astap=\aap
\let\apjlett=\apjl
\let\apjsupp=\apjs
\let\applopt=\ao

\newcommand{\ngc}{NGC~1624-2}
\newcommand{\hst}{\it HST\/}
\newcommand{\ha}{H$\alpha$}
\newcommand\ion[2] {#1\,{\sc{\romannumeral #2}}}  














\date{
Accepted 2018 November 23. Received 2018 November 21; in original form 2018 October 26}
\pagerange{\pageref{firstpage}--\pageref{lastpage}} \pubyear{2017}
\maketitle
\label{firstpage}

\begin{abstract}

In this paper, we present high-resolution HST/COS observations of the extreme magnetic O star NGC 1624-2. These represent the first 
ultraviolet spectra of this archetypal object. We examine the variability of its wind-sensitive resonance lines, comparing it to that of other known magnetic O stars. In particular, the observed variations in the profiles of the C\textsc{iv} and Si\textsc{iv} doublets between low state and high state are the largest observed in any magnetic O-type star, consistent with the expected properties of NGC 1624-2's magnetosphere. We also observe a redshifted absorption component in the low state, a feature not seen in most stars. We present preliminary modelling efforts based on the Analytic Dynamical Magnetosphere (ADM) formalism, demonstrating the necessity of using non-spherically symmetric models to determine wind/magnetospheric properties of magnetic O stars.

\end{abstract}



\section{Introduction}\label{sec:intro}

 
 

Recent spectropolarimetric surveys such as the Magnetism in Massive Stars (MiMeS; \citealt{2016MNRAS.456....2W, 2017MNRAS.465.2432G}) and the B-fields in OB stars (BOB; \citealt{2015IAUS..307..342M}) projects have led to the establishment of a distinct population of massive stars hosting detectable magnetic fields at their surfaces. About 7\% of Galactic OB stars are estimated to belong to this category. 

\citet{1972AJ.....77..312W} identified a peculiar spectroscopic class of O stars with strong emission in their C\textsc{iii} $\lambda$4650 line, comparable to that of the neighboring N\textsc{iii} lines, and labelled as ``Of?p". The most extreme example of this subclass is the O7f?p star NGC 1624-2 \citep{2010ApJ...711L.143W}. 
It has since been determined that this specific spectral feature (and various other observational properties of the Of?p stars) is an indication of magnetism \citep{2017MNRAS.465.2432G}. Indeed, 
all five Galactic Of?p stars host detectable fields, with NGC 1624-2 possessing the strongest known surface field of any O-type star by nearly an order of magnitude ($\sim 20$ kG; \citealt{2012MNRAS.425.1278W}). 

Since the magnetic and rotational axes of a massive star are (in general) not aligned, rotational modulations arise, explaining the variability observed for these stars throughout the electromagnetic spectrum. In this context, spectropolarimetric timeseries provide information on the topology of the field, as the line-of-sight magnetic configuration varies with phase since we view a different portion of the stellar surface as the star rotates. 
The Oblique Rotator Model (ORM; \citealt{1950MNRAS.110..395S}) describes this effect for a large-scale dipolar magnetic field. 

The H$\alpha$ emission of magnetic massive stars typically varies between a ``low state'' (corresponding to a nearly magnetic equator-on view) and a ``high state'' (pole-on view).
NGC 1624-2 exhibits a single-wave variation of the equivalent width of {\ha} over its rotational cycle \citep{2012MNRAS.425.1278W}, suggesting that we only view a single magnetic hemisphere, assuming a dipolar field (see, e.g., Fig. 1 from \citealt{2015MNRAS.453.3288P} for a schematic view of high and low state). 

In most magnetic massive stars, a number of other observables are found to vary over the rotational period, in a way that is consistent with the ORM paradigm, such as the profiles of wind-sensitive UV resonance lines. 
UV observations are particularly important to diagnose the wind properties of massive stars 
(e.g., \citealt{1994A&A...283..525P}), as wind-sensitive resonance lines contain information about their structure and kinematics
. UV properties have been studied for a few magnetic O stars: HD 37022 (= $\theta^1$ Ori C; \citealt{1996A&A...312..539S}), HD 57682 \citep{2009MNRAS.400L..94G,2012MNRAS.426.2208G}, HD 148937 \citep{2012A&A...538A..29M}, HD 108 \citep{2012MNRAS.422.2314M}, HD 191612 \citep{2013MNRAS.431.2253M}, CPD -28 2561 \citep{2015MNRAS.452.2641N} and HD 54879 \citep{2017A&A...606A..91S}.

The general phenomenology at UV wavelengths can be summarized as follows: at high state, strong lines show enhanced absorption at high velocities (compared to other rotational phases), while at low state, weaker lines exhibit strong low-velocity absorption (e.g., \citealt{2015MNRAS.452.2641N}). As described below, they also show a departure from the line profiles expected for a spherically symmetric outflow, e.g., the C\textsc{iv} $\lambda\lambda$1548/50 doublet is often found to be overall desaturated in the spectra of magnetic stars.

Magnetohydrodynamic (MHD) simulations have shown in great detail the process by which a magnetic field can confine the stellar wind into a circumstellar \textit{magnetosphere} \citep{2002ApJ...576..413U}. 
Slow rotators (such as NGC 1624-2, with a rotational period of 157.99d; \citealt{2012MNRAS.425.1278W}) form \textit{dynamical magnetospheres} (DMs), in which material flows along closed magnetic field loops from both magnetic hemispheres, 
colliding near the magnetic equator and then falling back onto the stellar surface, leading to complex and unstable flows.

Given its early spectral type, the dense stellar wind and extreme magnetic field of NGC 1624-2 
interact strongly, creating a giant magnetosphere, much larger and denser than that of any other OB star. This is evidenced 
by the strong emission observed in its H$\alpha$ line \citep{2012MNRAS.425.1278W}. 
In addition, the strong X-ray emission that is produced in magnetically confined wind shocks (MCWS) is substantially attenuated in NGC 1624-2, and varies with rotational phase \citep{2014ApJS..215...10N, 2015MNRAS.453.3288P}. 
Because of its extreme properties, NGC 1624-2 constitutes 
the ``Rosetta stone" of massive star magnetism
. Despite its importance, there have been no UV observations of this unique object until recently. In this paper, we 
present the first UV spectra of this object, obtained with the Cosmic Origins Spectrograph (COS) on the \textit{Hubble Space Telescope} (HST).

In Section~\ref{sec:obs} we discuss the observations. Section~\ref{sec:spec} offers a qualitative description of the spectra and compares NGC 1624-2 to other non-magnetic and magnetic stars of similar spectral type. In Section~\ref{sec:model} we describe our preliminary modelling methods and show some early results, including 
a comparison of NGC 1624-2's resonance line profiles to synthetic lines computed using the Analytic Dynamical Magnetosphere (ADM) formalism of \citet{2016MNRAS.462.3830O}. 
Finally, in Section~\ref{sec:concl} we draw conclusions and discuss future work.


\section{Observations}\label{sec:obs}



Ultraviolet spectra of {\ngc} were obtained at two rotational phases with the Cosmic Origins Spectrograph (COS) on board the {\it Hubble Space Telescope}. The spectra were acquired near phases corresponding to the ``high'' ($\phi \sim 0.0$) and ``low" ($\phi \sim 0.50$) states of the {\ha} variations. Table~{\ref{tab:obs}} provides a journal of these observations, which constitute {\hst} General Observer Program 13734 (PI: Petit).

For each visit, the data were collected with COS over two consecutive orbits. The first orbit was devoted to a target acquisition in the bright object aperture (BOA), which is required to place {\ngc} accurately in the circular primary science aperture (PSA; $2.5 \arcsec$ diameter).
The target acquisition was followed by two exposures with the G130M grating 
centered on wavelengths of 1291 and 1327~{\AA}, respectively.
During the second orbit, exposures with the G160M grating centered on
wavelengths of 1577 and 1623~{\AA} were obtained. These combinations provided complete wavelength coverage from about 1130 to 1800~{\AA} with a resolving power that increases linearly with wavelength from ~16,000 to ~21,000
over the respective ranges of both gratings. At each grating setting, the exposure was divided into 4 sub-exposures of equal duration but different FP-POS positions to mitigate the effects of fixed-pattern noise in the combined spectrum. All observations were obtained in time-tag mode by using both channels of the FUV cross-delay (XDL) line detector.

The spectra were uniformly processed with version 3.0 (2014 October 30) of the CALCOS calibration pipeline. The steps included: correcting the photon-event table for dead-time and positional effects such as drifts in the detector electronics, geometric distortion, and the Doppler shift of the observatory; binning the time-tag data and assigning wavelengths to the bins on the basis of Pt-Ne spectra acquired simultaneously with spectra of {\ngc}; extraction and photometric calibration of 1-D spectra; and the ``shift and add" combination of spectra taken at different FP-POS positions of individual grating settings. Since the first spectra of {\ngc} were obtained shortly after the move to life-time adjustment position 3 was implemented (on 2015 February 9), special care was taken to process the data with the appropriate reference files. Fiducial values of the signal-to-noise ratio (S/N) per 9-pixel resolution element are indicated in Table~\ref{tab:obs}.  These estimates were computed from the mean flux between 1350 and 1355~{\AA} and 1490 and 1495~{\AA} for the settings of the G130M and G160M gratings, respectively.  Since these intervals contain weak spectral features, the pixel-to-pixel noise was estimated from the scatter in the difference between the processed flux vector and a version that was smoothed over a resolution element.

Finally, the IDL procedure \texttt{coadd\char`_x1d}\footnote{This procedure was developed by the COS Instrument Definition Team.  See \citet{2010ApJ...720..976D} for a description and a hyperlink to the software repository.} was used to merge the extracted, calibrated spectra into a single spectrum.



\begin{table*}
\caption[c]{Journal of COS Observations}\label{tab:obs}
\begin{tabular}{@{}lccccccc}
\hline
ObsID     & Grating & {${\lambda_c}^1$} & S/N & UT (Start)         & Exp. Time & MJD(mid)   & $\phi\,^2$ \\
          &         &    (\AA)  &        &                    &  (s)      &            &            \\
\hline                 
lcl601010 &  G130M  &  1291 & 15.8           & 2015-02-17T06:16:42 &  ~848     & 57070.2685 & 0.986      \\
lcl601020 &  G130M  &  1327 & 14.8           & 2015-02-17T06:39:53 &  ~844     & 57070.2846 & 0.986      \\
lcl601030 &  G160M  &  1577 & 22.9           & 2015-02-17T07:40:05 &  1084     & 57070.3279 & 0.986      \\       
lcl601040 &  G160M  &  1623 & 22.6           & 2015-02-17T08:07:48 &  1080     & 57070.3472 & 0.986      \\   
lcl602010 &  G130M  &  1291 & 16.9           & 2015-10-06T16:20:53 &  ~848     & 57301.6881 & 0.451      \\  
lcl602020 &  G130M  &  1327 & 15.0           & 2015-10-06T16:44:04 &  ~844     & 57301.7042 & 0.451      \\ 
lcl602030 &  G160M  &  1577 & 21.6           & 2015-10-06T17:07:58 &  1084     & 57301.7309 & 0.451      \\
lcl602040 &  G160M  &  1623 & 21.1           & 2015-10-06T18:00:30 &  1080     & 57301.7588 & 0.452      \\       
\hline
\multicolumn{7}{l}{$^1$ Central wavelength of the grating setting.}                                    \\
\multicolumn{7}{l}{$^2$ Phase of MJD(mid) according to the ephemeris of \citealt{2012MNRAS.425.1278W}.}\\
\end{tabular}
\end{table*}

\section{Comparison to stars of similar spectral type}\label{sec:spec}

\begin{figure*}
\includegraphics[width=\textwidth]{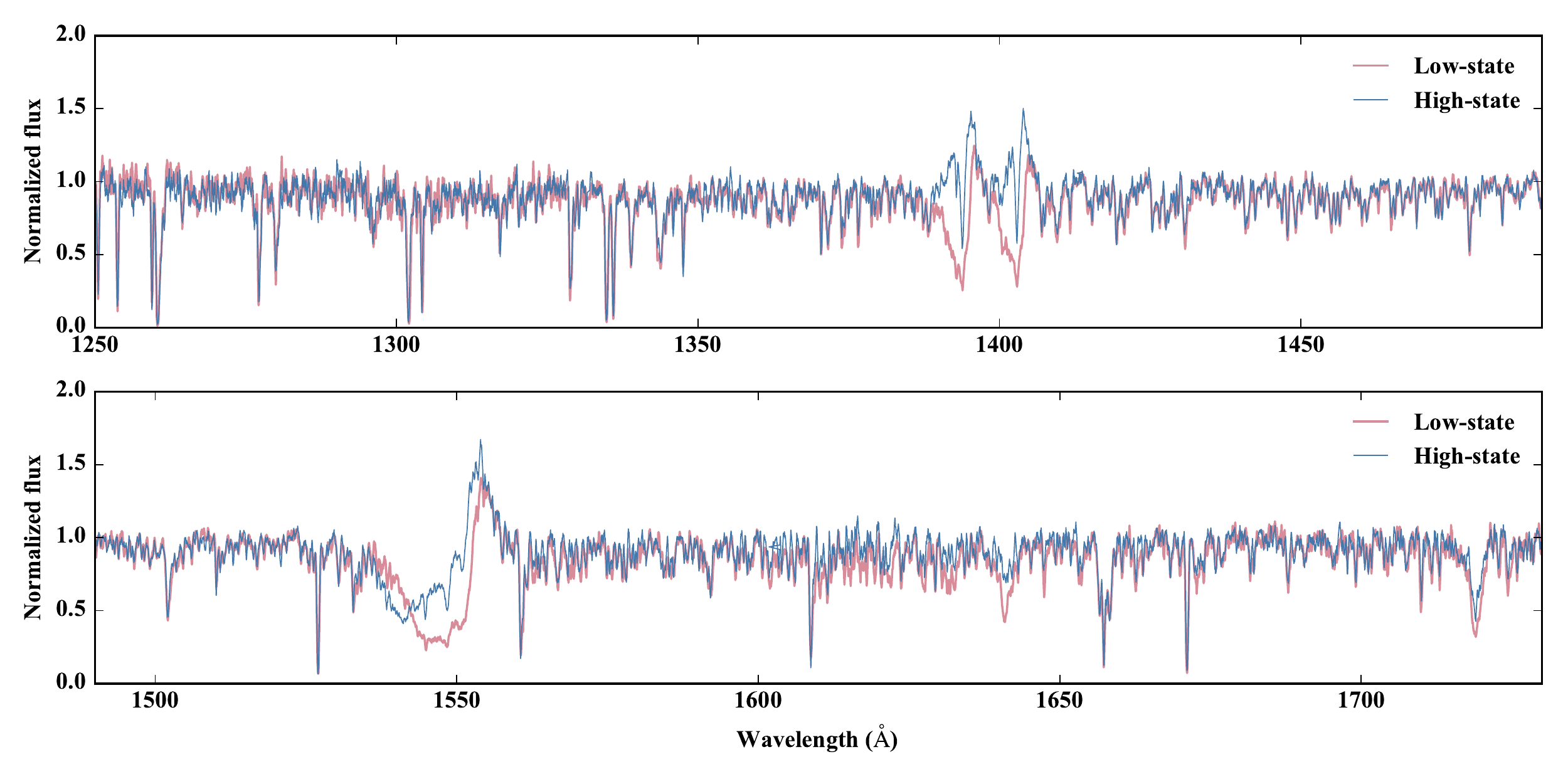}
\caption{\label{fig:compCMFGEN}\textit{HST} 
observations of NGC 1624-2 (high-state in blue
, low-state in pink) 
smoothed with an 11-pixel wide boxcar filter for display purposes. 
There are important 
variations in both the Si\textsc{iv}$\lambda\lambda$1393/1402 and C\textsc{iv}$\lambda\lambda$1548/50 doublets, as well as a modulation of the strength of the absorption in the Fe\textsc{iv} ``forest'' around 1615-1630\AA.
}
\end{figure*}

The UV spectra of NGC 1624-2 are shown in Figure~\ref{fig:compCMFGEN}. The low-state spectrum is shown in pink and the high-state spectrum is shown in blue.

\subsection{Qualitative description of the main UV spectral features of NGC 1624-2 and comparison with non-magnetic O stars}


 


\begin{figure*}
\includegraphics[width=\textwidth]{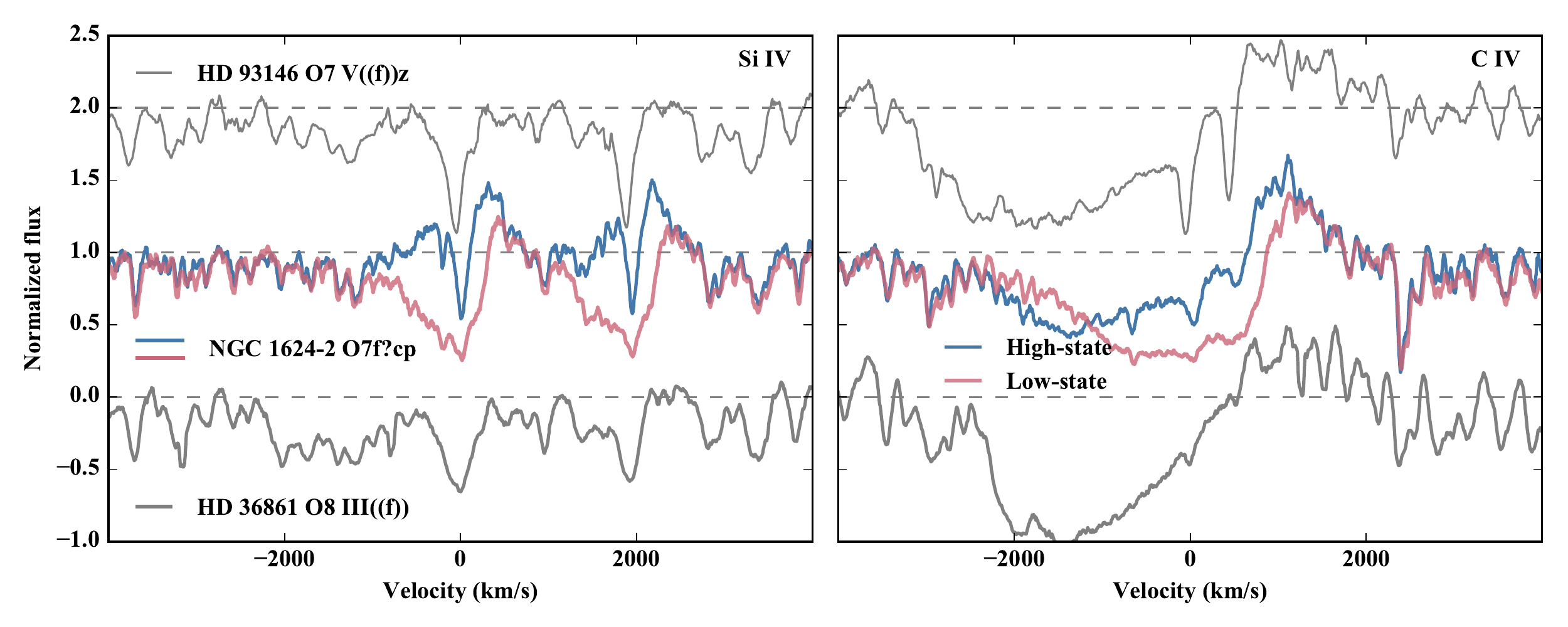}
\caption{\label{fig:compNormal} Comparison between the resonance line profiles of NGC 1624-2 and that of non-magnetic stars of similar spectral type (Si\textsc{iv}  $\lambda\lambda$1393/1402 on the left and C\textsc{iv} $\lambda\lambda$1548/50 on the right). The velocity axes refer to the blue component of the doublets. We see important differences in the morphologies of these lines, suggesting that the wind structure of NGC 1624-2 shows a strong departure from spherical symmetry. 
}
\end{figure*}


We first note that some of the most prominent spectral features undergo large variations over time. In particular, 
two important wind-sensitive resonance line doublets (Si\textsc{iv} $\lambda\lambda$1393/1402 and C\textsc{iv} $\lambda\lambda$1548/50) 
show clear differences in their 
line profiles between both phases due to rotational modulation as the viewing angle of the magnetosphere changes.

There is 
more absorption overall during the low state than during the high state. This is 
expected, since when the magnetosphere is viewed equator-on, there is a large amount of confined material obscuring the stellar disk
. 
However, despite the overall absorption being stronger, we can see that for 
the C\textsc{iv} doublet, 
the high-velocity, blue-shifted absorption is actually stronger during the high state. 
The emission peak of the doublet remains at roughly the same level at both phases for C\textsc{iv}, but in the case of Si\textsc{iv}, the overall emission is clearly stronger during the high state.

Fig. \ref{fig:compNormal} 
compares the profiles of NGC 1624-2's main wind-sensitive lines with those of two non-magnetic stars of similar spectral type: O7V((f))z (HD 93146, top) and O8III((f)) (HD 36861, bottom).

Comparing NGC 1624-2 with the dwarf O-type star HD 93146\footnote{Two components are known, HD 93146A and HD 93146B (O9.7IV; \citealt{2014ApJS..211...10S}), although it is not known whether they are physically bound. This might even be a hierarchical triple system \citep{2017A&A...600A..33M}, but in any case, the light is dominated by the primary. While the other components could fill the line profile slightly, contributing to its desaturation (in the case of the C\textsc{iv} doublet), this profile appears consistent with that of other stars of similar spectral type \citep{1985NASRP1155.....W}.}, we find that they both show similarly desaturated absorption troughs in their C\textsc{iv} doublet. The blue edge of the line profile of NGC 1624-2 appears at a lower velocity than that of the non-magnetic star, and in fact suggests a terminal velocity that is significantly lower than what is theoretically expected (2875 km s$^{-1}$; \citealt{2012MNRAS.425.1278W}). However the emission component of NGC 1624-2 is stronger.

On the other hand, while the absorption trough of the C\textsc{iv} doublet in the O giant HD 36861 is strongly saturated, its emission still is not as strong as that exhibited by NGC 1624-2.

Finally, when compared to both stars, NGC 1624-2's Si\textsc{iv} doublet stands out as it presents a fundamentally different morphology. In both non-magnetic stars, it is essentially unaffected by the wind, as is to be expected for a late O star. However, we can see a clear P Cygni-like profile in NGC 1624-2 
 during the low state. The double-peaked emission morphology present at the high state is to date unique to NGC 1624-2 and could arise from low-velocity material near the magnetic equator flowing from both magnetic hemispheres (therefore some material has a positive net line-of-sight velocity while the other half has a negative velocity), just off the limb of the star as it is viewed pole-on.


\subsection{Comparison with other magnetic O-type stars}



\begin{figure*}
\includegraphics[width=\textwidth]{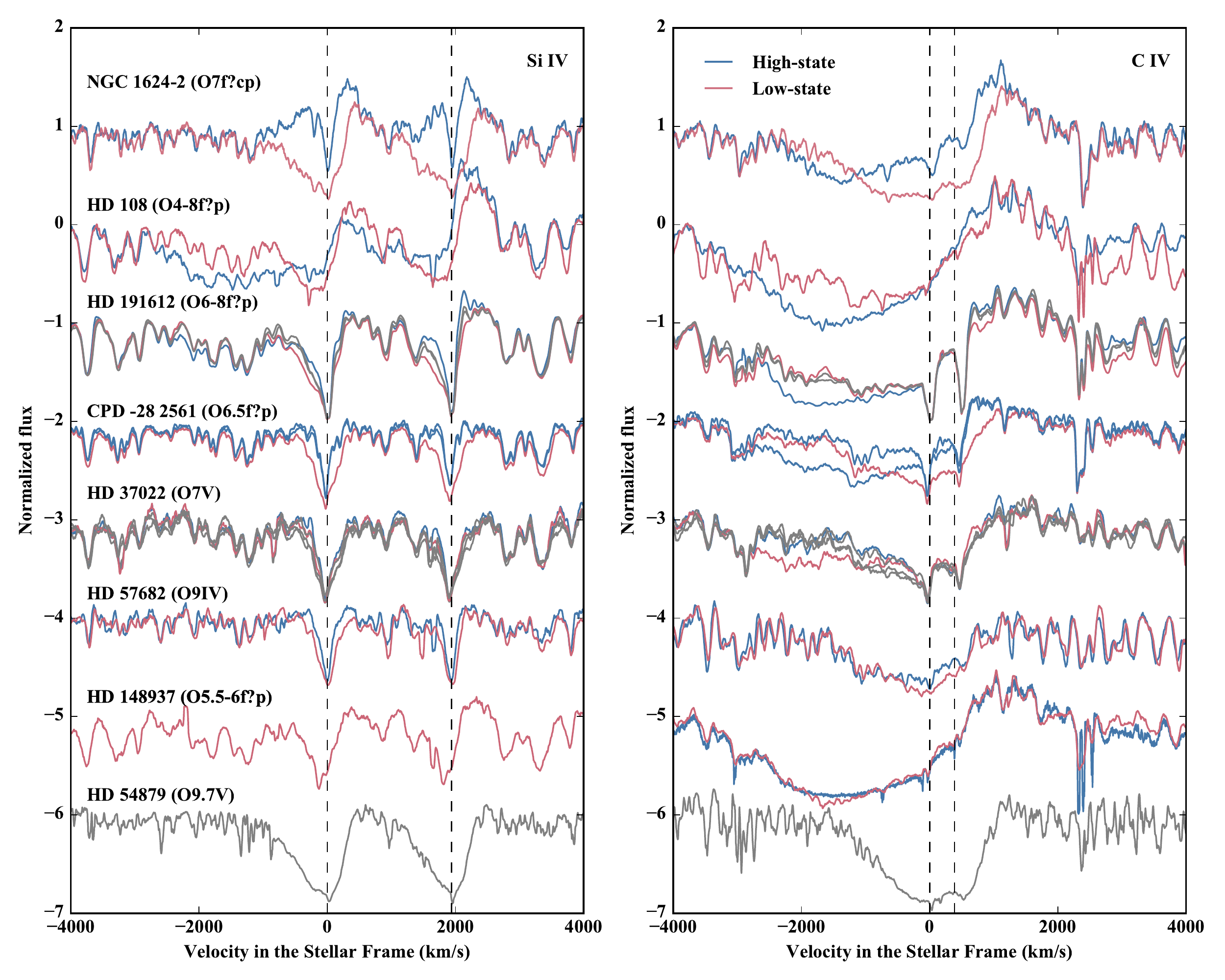}
\caption{\label{fig:compMag}Comparison between the resonance line profiles of NGC 1624-2 and that of other magnetic O-type stars. The velocity axis is centered on the position of the blue component of the doublet.
}
\end{figure*}

In Fig.~\ref{fig:compMag}, we visually compare the main wind-sensitive features of NGC 1624-2's spectra (namely the aforementioned Si\textsc{iv} and C\textsc{iv} doublets) to those of 7 other magnetic O stars for which archival UV spectra are available (though not always at extreme phases; blue lines denote high-state spectra and pink lines denote low-state spectra, while grey lines correspond to all other phases). 
The stellar and magnetospheric properties of these stars are summarized in Table~\ref{tab:mag}, and details regarding the observations can be found in the Appendix. 

The first obvious observation is that the resonance line profiles of NGC 1624-2 exhibit a much more pronounced variability than that of the comparison stars. This can undoubtedly be understood in terms of its larger and denser magnetosphere. Furthermore, as it is viewed nearly pole-on at high-state and nearly equator-on at low-state, we can probe roughly the full range of possible viewing angles. Its field configuration thus yields the greatest possible variability given its magnetospheric parameters, along with perhaps CPD -28 2561 and HD 57682 (for which both poles can presumably be seen over a rotational cycle, as suggested by their double-wave H$\alpha$ variations; see Table~\ref{tab:mag}). 


\begin{table*}
\caption[
Comparison magnetic O stars]{List of magnetic O-type stars appearing in Fig.~\ref{fig:compMag} with their stellar and magnetospheric properties: effective temperature ($T_\textrm{eff}$), mass ($M_*$), luminosity ($L$), mass-driving rate ($\dot{M}_{\textrm{B}=0}$), as calculated using the theoretical prescription by \citet{2001A&A...369..574V}, dipolar field strength ($B_p$), inclination angle ($i$; the angle between the rotation axis and the line of sight), the  obliquity angle ($\beta$; the angle between the magnetic field axis and the rotation axis) and the range of viewing angles ($\alpha$; the angle between the magnetic field axis and the line of sight). References for the stellar and magnetic parameters are the following: \textit{a})
\citet{2012MNRAS.425.1278W}; \textit{b}) \citet{2010MNRAS.407.1423M}; \textit{c}) \citet{2017MNRAS.468.3985S}; \textit{d}) \citet{2012A&A...538A..29M}; \textit{e}) \citet{2007MNRAS.381..433H}; \textit{f}) \citet{2011MNRAS.416.3160W}; \textit{g}) \citet{2015MNRAS.447.2551W}; \textit{h}) \citet{2010ApJ...711L.143W}; \textit{i}) \citet{2006A&A...448..351S}; \textit{j}) \citet{2006A&A...451..195W}; \textit{k}) \citet{2009MNRAS.400L..94G}; \textit{l}) \citet{2012MNRAS.426.2208G}; \textit{m}) \citet{2008AJ....135.1946N}; \textit{n}) \citet{2012MNRAS.419.2459W}; \textit{o}) \citet{2017A&A...606A..91S}; \textit{p}) \citet{2015A&A...581A..81C} and \textit{q}) \citet{2011ApJS..193...24S}.}\label{tab:mag}
\centering
\begin{tabular}{|c|c|c|c|c|c|c|c|c|}
	\hline
Star & $T_\textrm{eff}$ & $M_*$ & $\log(L)$ & $\log(\dot{M}_{\textrm{B}=0})$ & $B_p$ & $i$ & $\beta$ & $\alpha$ range\\
Spectral type     & (kK)        &    $(\textrm{M}_\odot)$ & $(\textrm{L}_\odot)$     &      $(\textrm{M}_\odot \textrm{yr}^{-1})$     & (kG) & ($^{\circ}$) & ($^{\circ}$) & ($^{\circ}$) \\
	\hline
NGC 1624-2 & $35\pm 2$ $^{a}$ & $\sim 30$ $^{a}$ & $5.2 \pm 0.1$ $^{a}$ & $-6.8$ & $\sim 20$ $^{a}$ & $\lesssim 45$ $^{a}$ & $\lesssim 45$ $^{a}$ & $\sim 0-90$ \\
O7f?cp $^{a}$ & & & & & & & & \\
\hline
HD 108   & $35\pm 2$ $^{b}$ & $42 \pm 5$ $^{c}$ & $5.7 \pm 0.1$ $^{b}$ & $-5.55$ & $\geq 1.15$ $^{c}$ & N/A & N/A & N/A \\
O4-8f?p $^{d}$ & & & & & & & & \\
\hline
HD 191612 & $35 \pm 1$ $^{e}$ & $\sim 30$ $^{e}$ & $\sim 5.4$ $^{e}$ & $\sim -6.0$ & $2.45 \pm 0.40$ $^{f}$ & $\sim 30$ $^{f}$ & $67 \pm 5$ $^{f}$ & $\sim 37-97$\\
O6-8f?p $^{d}$ & & & & & & & & \\
\hline
CPD -28 2561 & $35 \pm 2$ $^{g}$ & $61 \pm 33$ $^{g}$ & $5.35 \pm 0.15$ $^{g}$ & $-6.56$ & $2.6 \pm 0.9$ $^{g}$ & $35 \pm 3$ $^{g}$ & $90 \pm 4$ $^{g}$ & $\sim 55-125$ \\
O6.5f?p $^{h}$ & & & & & & & & \\
\hline
HD 37022 & $39 \pm 1$ $^{i}$ & $45 \pm 16$ $^{i}$ & $5.31 \pm 0.13$ $^{i}$ & $-6.40$ & $1.525 \pm 0.375$ $^{j}$ & $45 \pm 20$ $^{j}$ & $\sim 48 \pm 20$ $^{j}$ & $\sim 3-93$\\
O7V $^{d}$ & & & & & & & & \\
\hline
HD 57682 & $34.5 \pm 1.0$ $^{k}$ & $17^{+19}_{-9}$ $^{k}$ & $4.79 \pm 0.25$ $^{k}$ & $-7.08$ & $0.88 \pm 0.05$ $^{l}$ & $\sim 60$ $^{l}$ & $79 \pm 4$ $^{l}$ & $\sim 19-139$\\
O9IV $^{d}$ & & & & & & & & \\
\hline
HD 148937 & $41 \pm 2$ $^{m}$ & $\sim 60$ $^{n}$ & $5.8 \pm 0.1$ $^{n}$ & $\sim -5.5$ & $1.02^{+0.31}_{-0.38}$ $^{n}$ & $\lesssim 30$ $^{n}$ & $38^{+17}_{-28}$ $^{n}$ & $\sim 8-68$\\
O5.5-6f?p $^{d}$ & & & & & & & & \\
\hline
HD 54879 & $30.5 \pm 0.5$ $^{o}$ & $14 \pm 7$ $^{o}$ & $4.45 \pm 0.20$ $^{o}$ & $-7.9$ & $\geq 2$ $^{p}$ & N/A & N/A & N/A \\
O9.7V $^{q}$ & & & & & & & & \\
\hline
\end{tabular}
\end{table*}


At its low state, NGC 1624-2's Si\textsc{iv} doublet 
shows a clear P-Cygni profile which is more reminiscent of an early-type O giant, or an OB supergiant. 
Among other magnetic O stars, only HD 108 shows a similar emission. However, it does not share the remarkable double-peaked emission morphology that NGC 1624-2 exhibits in its high state. 
The other stars mostly exhibit minor variations of this doublet, with a slight broadening and more absorption at low state. 

In contrast, the C\textsc{iv} doublet of NGC 1624-2 
shows less absorption at high velocity during the low state, but more absorption at low velocity (which also manifests as an apparent decrease of the red-shifted emission). This is consistent with synthetic lines computed from MHD simulations \citep{2013MNRAS.431.2253M}, and similar to the variation seen in CPD -28 2561.

Another interesting feature of NGC 1624-2's UV resonance lines is 
the presence of broad, red-shifted absorption at low-state in both doublets. Among the other magnetic O stars, this feature is only seen in HD 54879, 
and could be linked to infalling material. 
In fact, HD 54879's UV spectrum is surprisingly similar to NGC 1624-2's low-state spectrum, despite its later spectral type.

Finally, we can also see from Fig.~\ref{fig:compCMFGEN} that the absorption in the ``forest'' of Fe\textsc{iv} photospheric lines is modulated between phases, appearing to be stronger at low state. This behaviour was also observed in the cases of HD 108 and HD 191612 \citep{2013MNRAS.431.2253M}.


As was pointed out by \citet{2015MNRAS.452.2641N}, comparing the behaviour of the UV resonance lines of all observed magnetic O 
stars (with the exception of HD 37022, which puzzlingly seems to exhibit the exact opposite behaviour; see section~\ref{subs:prev}) 
leads to an overall consistent phenomenology for the variations seen in the UV wind-sensitive lines of magnetic O stars. In the case of \textit{strong}\footnote{In other words, lines which correspond to electronic transitions with a large opacity.} lines, which can probe low-density, high-velocity material further from the stellar surface, the P-Cygni characteristics of the line profile seem to be intensified at high state. On the other hand, \textit{weak} lines are formed in regions of high density, that can provide enough optical depth despite their weak opacity; these high-density regions are usually found surrounding the magnetic equator, closer to the star and therefore contribute significant low-velocity absorption at low state. 

This means that there are two main factors in understanding the nature and amplitude of the line profile variations detected in UV lines: line strength and the range of magnetospheric viewing angles. As further discussed in the following section, the line strength itself depends sensitively on the ionization balance and the mass-loss rate (or in the case of magnetic massive stars, the wind-feeding rate as discussed below
). This can explain the range of behaviours and line morphologies examined in this subsection. As for the viewing angle, it appears that the extremes of these line profile variations 
correspond to a pole-on view ($\alpha = 0^{\circ}$) and an equator-on view ($\alpha = 90^{\circ}$), so a larger range in viewing angle between those two values will lead to larger variations. Conversely, for HD 148937, the limited range in viewing angle precludes sampling the full range of variations. Finally, depending on how far from the surface a wind-sensitive line is formed, it might be more or less affected by magnetic wind confinement depending on the Alfv\'{e}n radius.

\section{Modelling}\label{sec:model}



\subsection{Previous modelling efforts}\label{subs:prev}

In non-magnetic stars, UV resonance lines provide a critical diagnosis of wind properties. Spectral synthesis codes such as \textsc{cmfgen} \citep{1998ApJ...496..407H} use models with spherically symmetric winds. Therefore, fitting the UV spectrum of a non-magnetic massive star with one of these models yields the mass-loss rate and terminal velocity of the wind. 

However, the picture is more complicated in a magnetic massive star, since the field impacts both the density and velocity distribution of the circumstellar 
material, reshaping the wind into a magnetosphere.
The usual 
assumption to circumvent this issue has been to consider, whenever possible, the high state spectrum to perform a quantitative spectral analysis using spherically symmetric models
. Close to the magnetic pole the wind breaks open the field lines and the flow is nearly radial; therefore this viewing angle shows us the part of the magnetosphere that 
most resembles a spherically symmetric wind. 

Such an approach has been adopted using \textsc{cmfgen} in the cases of HD 108 \citep{2012MNRAS.422.2314M}, HD 191612 \citep{2013MNRAS.431.2253M} and HD 57682 \citep{2009MNRAS.400L..94G}, and using the PoWR code \citep{2002A&A...387..244G, 2003A&A...410..993H} in the case of HD 54879 \citep{2017A&A...606A..91S}. However, these attempts 
were typically 
not able to account for the detailed shapes of the line profiles and yielded mass-loss rates that are at least an order of magnitude lower than the expected theoretical rates. In the case of HD 57682, \citet{2012MNRAS.426.2208G} used a heuristic model with a disc of circumstellar material concentrated around the magnetic equator to reproduce the observed variations in the H$\alpha$ emission. This model yielded a mass-loss rate which was between 10 and 30 times greater than that inferred from the previous \textsc{cmfgen} fit to the UV spectrum.

Spherically symmetric models can also be used to determine the ionization balance in the wind. This approach was used for instance by \citet{2015MNRAS.452.2641N} for CPD -28 2561, using \textsc{fastwind} (\citealt{2005A&A...435..669P}, combined with a new X-ray treatment implemented by \citealt{2016A&A...590A..88C}).

It has been shown that the behaviour of the UV resonance lines of magnetic O stars can be qualitatively understood by performing radiative transfer through MHD simulations. This technique has proven successful for both HD 191612 \citep{2013MNRAS.431.2253M} and CPD -28 2561 \citep{2015MNRAS.452.2641N}. In the former case, the opposite behaviours of the C\textsc{iv} and Si\textsc{iv} doublets is properly reproduced. For the latter star, MHD simulations coupled with a \textsc{tlusty} photospheric profile \citep{2003ApJS..146..417L} and radiative transfer can reproduce the observed broadening of the Si\textsc{iv} doublet's absorption at low state.

However, MHD simulations fail to reproduce the observed variability of HD 37022, showing instead an offset of 0.5 in rotational phase \citep{2008cihw.conf..125U}. While this could perhaps be understood in terms of certain hypotheses formulated by \citet{1996A&A...312..539S} regarding the magnetic field topology (well before the field was even detected), further magnetic characterization of this star by \citet{2006A&A...451..195W} conclusively excludes this possibility, making this phenomenology an enduring puzzle.

\subsection{Departure from spherically symmetric winds}\label{sec:depssw}





\begin{figure*}
\includegraphics[width=\textwidth]{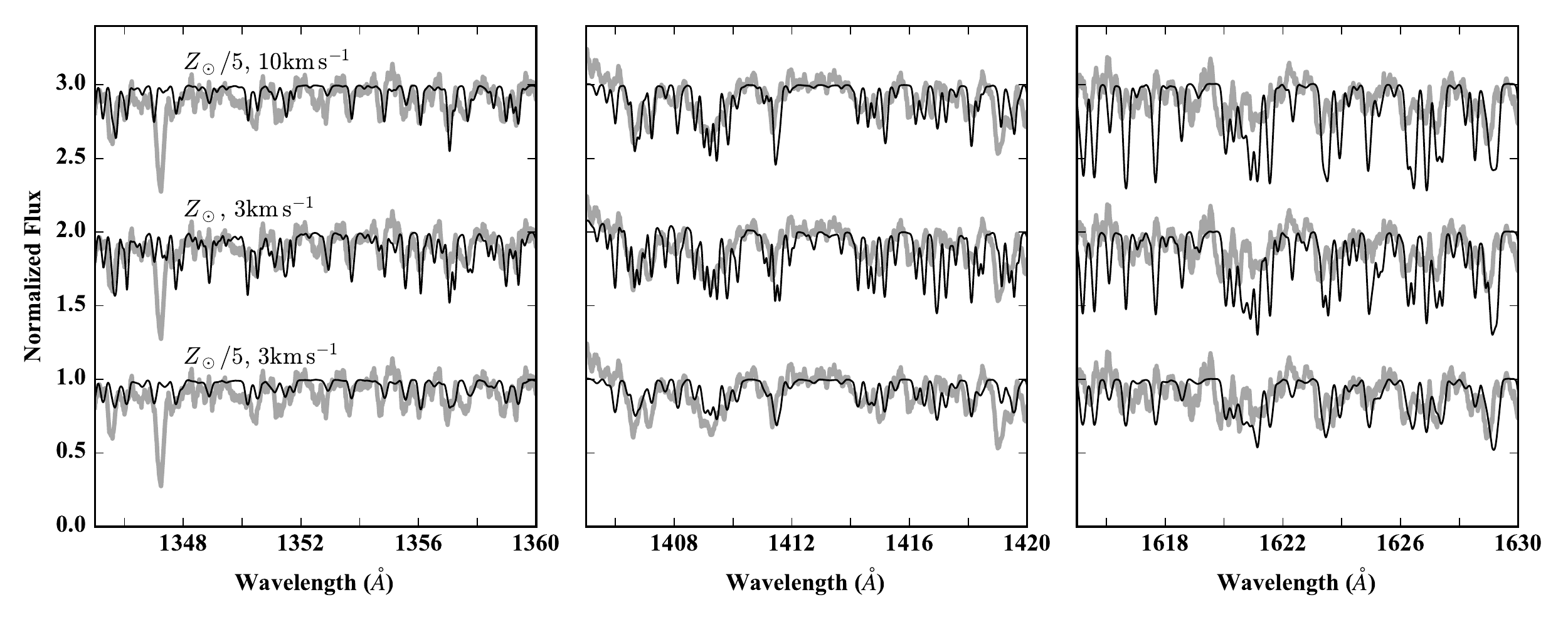}
\caption{\label{fig:feforest}Fe\textsc{v} (left and middle panel) and Fe\textsc{iv} (right panel) ``forests'' of photospheric lines. The high-state observation is represented by the thick grey lines and the models by the thin black lines. The observation is compared to the best fit \textsc{cmfgen} model on the bottom, which uses both subsolar metallicity and low microturbulent velocity, and to two other models which implement each of these factors separately in the middle and on the top (vertically offset by 1 and 2, respectively). Only the model on the bottom can simultaneously match the line strengths for all three ``forests''.}
\end{figure*}

\citet{2012MNRAS.425.1278W} used the optical spectrum of NGC 1624-2 to derive its stellar properties, and then computed a mass-loss rate of $1.6 \times 10^{-7} \textrm{M}_\odot \textrm{yr}^{-1}$ using the prescription by \citet{2001A&A...369..574V}. However, it is important to note that this corresponds to the value the mass-loss rate would have if the star did not possess a magnetic field, or in other words the \textit{wind-feeding rate, $\dot{M}_{B=0}$}. The parameterizations derived by, e.g., \citet{2002ApJ...576..413U,2008MNRAS.385...97U,2009MNRAS.392.1022U} to describe various quantities related to massive star magnetospheres (such as the confinement parameter, spin-down time and Alfv\'{e}n radius) are also all expressed in terms of the wind-feeding rate.

In reality, however, this wind-feeding rate does not correspond to the total surface mass flux, since the local mass flux varies latitudinally due to the inclination of the magnetic field with respect to the normal to the stellar surface 
\citep{2004ApJ...600.1004O}. This leads to a reduction of the integrated surface mass flux by a factor of nearly one-half in the case of a pure dipole \citep{2016MNRAS.462.3830O}. Furthermore, the magnetic field confines a large portion of the outflowing material (up to 96\% in the case of NGC 1624-2), which means that the ``real'' mass-loss rate should only take into account the material that actually escapes the star, mostly the outflow around the polar region that is not confined (e.g., \citealt{2008MNRAS.385...97U, 2017MNRAS.466.1052P}). This ``real'' mass-loss rate is of particular interest in the context of stellar evolution modelling (e.g., \citealt{2017MNRAS.466.1052P, 2017IAUS..329..250K, 2017A&A...599L...5G}).

To recapitulate, we are distinguishing between three main quantities:

\begin{enumerate}
\item the ``wind-feeding rate'' (or $\dot{M}_{B=0}$) which corresponds to the expected mass-loss rate (in our case obtained via a theoretical prescription) a non-magnetic star with the same stellar parameters would have;
\item the integrated surface mass flux, which, compared to the previous quantity, is reduced by some factor due to the inclination of the surface magnetic field lines; and
\item the ``real'' mass-loss rate, which corresponds to the material escaping both the star's gravity and the magnetic confinement along open field lines.
\end{enumerate}

The question then becomes which one (if any) of these rates would UV line fitting 
yield? Ideally, we would like to use a model that, taking into account the magnetospheric geometry, would recover $\dot{M}_{B=0}$. First, we consider, as has been done traditionally
, the case of a spherically symmetric model. 


We compute 
\textsc{cmfgen} models using the unclumped wind-feeding rate ($\log \dot{M}_{B=0} = -6.8$) adopted by \citet{2012MNRAS.425.1278W} as well as an equivalent clumped wind-feeding rate of $\log \dot{M}_{B=0} = -7.3$ using a typical (e.g., \citealt{2002ApJ...581..258F}) volume-filling clumping factor of $f = 0.1$ (since $\dot{M} \propto \sqrt{f}$ for optically thin clumping). We also take into account the X-ray luminosity\footnote{Based on a fit to an ISM-corrected model for NGC 1624-2 in its high state.} ($L_\textrm{X}/L_\textrm{bol} = -6.2$) obtained by \citet{2015MNRAS.453.3288P}. Finally, we 
compare these models to our high-state spectrum. 
The agreement 
with the wind features is poor (see section~\ref{sec:results}), as expected, and we find that any attempt to fit the blue-shifted absorption of UV resonance lines would yield a wind-feeding rate at least one order of magnitude lower than the theoretical value (as has been found for other stars). This is due in part to the faster-than-radial expansion of a radial polar flow tube in a magnetic wind (e.g., \citealt{1976SoPh...49...43K, 2004ApJ...600.1004O}), as discussed by \citet{2013MNRAS.431.2253M}. 

However, if we focus on photospheric lines, such as the Fe\textsc{v} and Fe\textsc{iv} ``forests'', we notice some interesting features. In Fig.~\ref{fig:feforest} we show the associated regions in the (high-state) spectrum. 
Thin black lines are models for various combinations of microturbulent velocity ($v_{\textrm{turb}}$) and metallicity (and thus iron content). The effective temperature and surface gravity are those determined by \cite{2012MNRAS.425.1278W} based on the analysis of the optical spectrum. 
A standard model with solar composition ($Z_\odot$) and a mictroturbulent velocity of 10 km/s produces too much Fe\textsc{iv} absorption. Reducing $v_{\textrm{turb}}$ to 3 km/s (middle) improves the fit but the Fe\textsc{iv} absorption remains too strong. Further reducing $v_{\textrm{turb}}$ to 1 km/s (not shown) does not change this picture significantly. 
On the other hand, reducing only the metallicity 
with a fixed microturbulent velocity of 10 km/s (top) 
does not reduce the strength of the Fe\textsc{iv} absorption sufficiently to reproduce the 
observed spectrum. 
However, a model with 
a metallicity of $Z_\odot/5$ 
and with $v_{\textrm{turb}}$ = 3 km/s (bottom) simultaneously reproduces the absorption strength of both the Fe\textsc{v} and Fe\textsc{iv} ``forests''. Such a fit could not be accomplished by changing, e.g., the surface temperature, since that would also affect the depth of the Fe\textsc{v} ``forest'', thus simply shifting the problem from one set of lines to the other. We note that \cite{2013MNRAS.433.2497S} reported the suppression 
of 
turbulent broadening in NGC 1624-2 as observed in the optical spectra. 

Taking into account a lower metallicity should lead to a lower theoretical wind-feeding rate (by roughly half an order of magnitude). However, while this result is intriguing, a proper abundance analysis for a magnetic O star 
would benefit from using non-spherically-symmetric atmospheric models (as evidenced by, e.g., the modulation of the Fe\textsc{iv} forest between high and low state). Since such models are not currently available, we limit ourselves to noting the possibility of a subsolar metallicity\footnote{There have been no metallicity measurements for the cluster NGC 1624, but its position at the edge of the Galactic disk could be consistent with a subsolar value.}, while still using the theoretical mass-loss rate computed at solar metallicity by \citet{2012MNRAS.425.1278W} for our subsequent modelling and analysis.

\subsection{Magnetospheric modelling}


While it is possible to use MHD simulations to model massive star magnetospheres, 
they involve expensive calculations. Therefore, they are not particularly well suited to model a set of magnetic stars with various stellar and magnetic parameters. Furthermore, 
such calculations are not practical for a star like NGC 1624-2, due to its extremely strong magnetic field that leads to a high Alfv\'{e}n speed, and consequently to extremely short Courant stepping times. In contrast, the simplified parameterizations included in the ADM formalism provide a significantly more efficient method to model 
the time-averaged behaviour of this star's 
magnetosphere 
and might allow us to recover its wind-feeding rate.
This method has already been shown to successfully reproduce the H$\alpha$ variations of slowly rotating magnetic massive stars \citep{2016MNRAS.462.3830O}.

The ADM model consists of three main components. The \textit{upflow} is the ionized material radiatively driven from the surface of the star. The ADM formalism makes the assumption that the magnitude of the material's velocity follows a $\beta$-law\footnote{And in particular in this case, a velocity law using $\beta = 1$ such that the radial dependence of the velocity follows $v(r) = v_\infty (1 - R_*/r)$.}, but that its direction follows that of the field lines, as the material flows along the closed field loops from both magnetic hemispheres. 
This upflow leads to a region of shocked plasma, located between the shock front and the apex of the magnetic loop, close to the plane of the magnetic equator leading to a region of \textit{hot post-shock gas}, whose extent depends on the cooling efficiency. Finally, once cooled, the material simply forms a \textit{downflow} that starts at the loop apex and flows back onto the surface of the star along the field loops, undergoing free fall. Of course, the simultaneous presence of these three components at a given point in space is not physical, but it represents an adequate time-averaged picture of the dynamic flows which occur within the magnetosphere. Given these three components, the ADM model provides us with analytical prescriptions for the density and velocity at each point within the magnetosphere. 

To model NGC 1624-2's magnetosphere, we make a few simplifying assumptions and approximations. We first consider a pure dipole model with an infinite Alfv\'{e}n radius\footnote{In a dipole model, the Alfv\'{e}n radius corresponds roughly to the radius of closure of the largest closed field loop.} ($R_\textrm{A}$). 
In reality, for NGC 1624-2, $R_\textrm{A} \approx 11.4 R_*$, which is large enough for this approximation to be warranted, since the extremely low densities at that distance will not contribute much to the line of sight optical depth. 
The extent of the closed field loops in NGC 1624-2's magnetosphere also eliminates concerns 
that the ADM poorly reproduces the outer wind \citep{2018A&A...616A.140H}\footnote{This particular study used HD 191612, with a much smaller Alfv\'{e}n radius of $2.7 R_*$, to compare MHD simulations and the ADM formalism using, in effect, an infinite Alfv\'{e}n radius.}. 

Furthermore, for this initial investigation, we only consider the upflow component. This is because we are seeking to investigate the desaturation that occurs within the high-velocity absorption trough in the high state.
Since the downflow component located between the stellar disk and the observer (looking pole-on) would be redshifted, and the shock-heated region also should not cross the stellar disk from the observer's perspective (unless the cooling is unrealistically inefficient), using only the upflow provides a proof of concept to test whether the ADM formalism offers a better basis than previous methods to provide a quantitative estimate of the wind-feeding rate. 



\subsection{Radiative transfer}


To perform radiative transfer on our model magnetosphere, we use a variation of the Sobolev with Exact Integration (SEI) method  for a singlet \citep{1981A&A....93..353H, 1987ApJ...314..726L}, taking into account the $\kappa_{0}$ formalism of line strength as presented by \citet{2014A&A...568A..59S} (their eq. 13). This parameter is proportional to the mass-loss rate and the fractional abundance of the absorbing ion for a given line.

The scalings used in the ADM model take into account two important velocities: the terminal velocity of the wind, $v_\infty$, and the escape velocity,

\begin{equation}
v_e \equiv \sqrt{\frac{2 G M_* (1-\Gamma_e)}{R_*}}
\end{equation}

\noindent where $G$ is the universal gravitational constant, $M_*$ and $R_*$ are the stellar mass and radius, respectively, and

\begin{equation}
\Gamma_e \equiv \frac{\kappa_e L}{4 \pi G M_* c}
\end{equation}

\noindent is the standard Eddington factor, with $\kappa_e = 0.34 \ \textrm{cm}^2 \textrm{g}^{-1}$, the Thomson electron scattering opacity and $c$, the speed of light. Finally, we define

\begin{equation}
\gamma_w \equiv v_\infty / v_e
\end{equation}

\noindent to be the ratio between the terminal and escape velocities. Generally, for stars on the hot side of the bi-stability jump ($T_\textrm{eff} \gtrsim 21$ kK), it is assumed that $\gamma_w = 2.6$ \citep{1995ApJ...455..269L}.

Then, for a given frequency within the line profile $w_\textrm{obs}$, expressed in terms of velocity (normalized to $v_\infty$), we obtain the following expression for an infinitesimal element of optical depth for a point along the line-of-sight ($z$) direction:

\begin{equation}
d\tau(w_\textrm{obs},z) = \frac{\kappa_{0} \gamma_w}{\sqrt{\pi} w_D} \frac{\rho}{\rho_{c*}} \exp{\left[-\left(\frac{w_z-w_\textrm{obs}}{w_D}\right)^2\right]} dz
\end{equation}

\noindent where $w_D$ is a velocity broadening parameter taking into account both thermal and turbulent velocity fields (in units of $v_\infty$; for our models we adopt a value of 0.01, which correspond to a thermal broadening of 20-30 km/s), $\rho/\rho_{c*}$ is an output of the ADM model corresponding to the scaled density (where the scaling factor is $\rho_{c*} = \dot{M}_{\textrm{B}=0}/4\pi R^2_* v_e$), and $w_z$ is the line-of-sight component of the velocity (in units of $v_\infty$). The ADM model actually uses two different density scalings for the upflow and downflow components ($\rho_{c*}$ and $\rho_{w*}$), but they are easily related by using the $\gamma_w$ parameter, which for now is set at the canonical value of 2.6. 


Furthermore, given the scope of this study, we limit our investigation to the absorption due to the intervening material which is in front of the stellar disk, along the line of sight. This is suitable to model the high velocity absorption trough, which is nearly unaffected by light that is scattered back into the line of sight. Future efforts will, however, take into account the full radiative transfer solution. For that same reason, we do not need to take into account a base photospheric absorption profile.


From the perspective of radiative transfer, using an analytic prescription to model the magnetospheric structure yields the marked advantage of allowing us to preset our integration grid, and then in a single loop both calculate the model at those points and integrate over each ray. This leads to a very efficient line synthesis technique, ideal for a vast parameter study.

We compute the absorption along 317 rays covering the stellar disk, where each ray contains 1000 linearly spaced points along the line of sight. 
The optical depth is computed along each ray using a piecewise linear integration scheme.


\subsection{Modelling results}\label{sec:results}


First, we illustrate the strong desaturation that occurs at high velocity at a given line strength. We computed the optical depth for each of the rays for a specific frequency and created surface maps for two specific cases: a spherically symmetric wind, and a wind threaded by a pure dipole seen pole-on. The result can be seen in Fig.~\ref{fig4}. We chose a value of $w_{obs} = -0.8$, in the high-velocity absorption trough of the line. The optical depth is normalized to $\kappa_0$. We can see that the magnetized wind has a much weaker optical depth at high velocity compared to the radially outflowing one. This corresponds to an effect that has already been observed for magnetic massive stars (see Fig.~\ref{fig:compMag}).


\begin{figure}
\centering
\includegraphics[width=\linewidth]{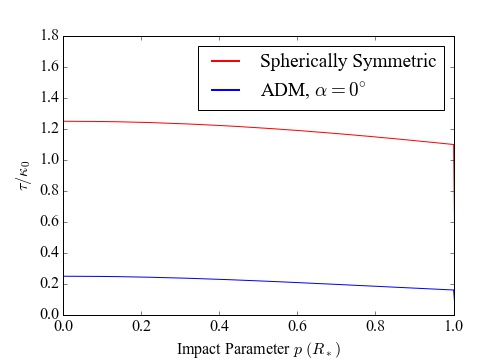} 
\label{fig4} 
\caption{Optical depth at the stellar surface (scaled to the line strength, $\kappa_0$) as a function of the impact parameter $p$ (scaled to the stellar radius, $R_*$) 
at $v/v_{\infty}=-0.8$, comparing a spherically symmetric outflowing stellar wind (red) with material trapped by a dipolar magnetic field viewed along the magnetic pole (blue). 
The decrease in optical depth between the spherically symmetric wind and the magnetized wind is immediately apparent, demonstrating that the presence of a magnetic field leads to the desaturation of the absorption trough observed in NGC 1624-2. 
 }
\label{fig4} 
\end{figure}

While this further demonstrates 
that spherically symmetric models are not suitable to infer wind properties from magnetic massive stars, it remains to be seen whether the ADM model can provide a quantitative approach to measure the wind-feeding rate of these objects. As a first proof of concept, we compute absorption-only line profiles and compare them to the pole-on spectrum of NGC 1624-2 and the \textsc{cmfgen} model shown in Fig.~\ref{fig:compCMFGEN}. We model the strong C\textsc{iv} doublet, which is typically used to diagnose wind parameters.


To do so, we must first compute the line strength ($\kappa_0$) associated to the computed theoretical wind-feeding rate of NGC 1624-2. As in \citet{2012MNRAS.425.1278W}, we use a wind-feeding rate of $\log(\dot{M}_{B=0}) = -6.8$. Since the ADM model does not provide any information regarding ionization states throughout the magnetosphere, 
we have to resort to using the ion fraction outputted by the \textsc{cmfgen} model, in this case $\log(n_{\textrm{C\textsc{IV}}}/n_\textrm{H}) \sim -5$. This leads to a value of $\kappa_0 = 5$.

Then, we validated this line strength value by comparing a profile obtained for a spherically symmetric outflow to the \textsc{cmfgen} model presented in Section~\ref{sec:depssw}. We computed profiles both with and without emission to determine at what velocity emission starts contributing significantly to the line profile; we found its contribution to be negligible for velocities ranging from $w_z = -1.0$ to $w_z = -0.8$. 

\begin{figure*}
\centering
\includegraphics[width=\linewidth]{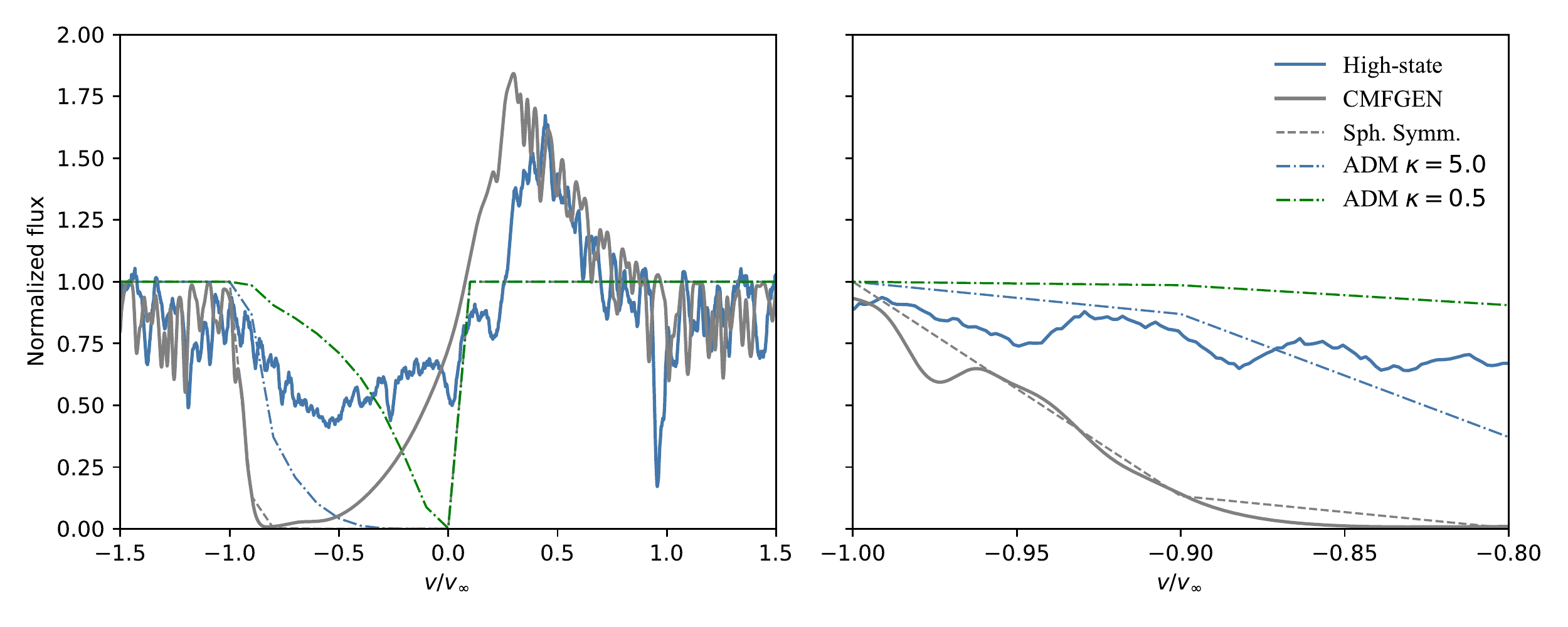} 
\caption{Comparison between the line profiles of the C\textsc{iv} doublet of NGC 1624-2 (high state, in solid blue), the \textsc{cmfgen} model (in solid grey) computed using the theoretical mass-loss rate derived by \citet{2012MNRAS.425.1278W} (see Section~\ref{sec:depssw}), and our absorption-only singlet model calculations using both a spherically symmetric wind (in dashed grey) and the ADM prescription (for line strengths corresponding to the theoretical wind-feeding rate in dash-dotted blue, and a wind-feeding rate which is one order of magnitude smaller in dash-dotted green). All profiles are normalized to their respective empirical or theoretical terminal velocity. The left panel shows the full line profiles, while the right panel focuses on the high-velocity blue-shifted absorption trough. As expected, there is good agreement between the \textsc{cmfgen} model and our spherically symmetric model. In addition, the ADM profile with a line strength of $\kappa_0 = 5$ closely follows the observations of NGC 1624-2 in this regime (the departure at lower velocities is due to the missing emission in our model), while the ADM profile with a weaker line strength does not produce enough absorption to match the observation, even at lower velocities despite the lack of emission in the model.} 
\label{fig:lines}
\end{figure*}

We finally used the same line strength to compute a line profile (absorption only) using the ADM model viewed pole-on ($\alpha = 0^{\circ}$). 
The results are shown in Fig.~\ref{fig:lines}. We can see that at high blue-shifted velocities (beyond $w_z = -0.8$), our spherically symmetric model is consistent with the \textsc{cmfgen} model. Interestingly, the ADM model that we computed also seems to reproduce the high velocity line depth of the actual pole-on data, at least beyond $w_z = -0.9$. The discrepancy which arises at lower velocities obviously stems from the lack of emission in our profile, but could also be due to the magnetic geometry of NGC 1624-2: if the high state corresponds to a slightly off-pole view, this would lead to an even shallower absorption trough. For comparison, we also computed another ADM model line profile with a smaller line strength of $\kappa_0 = 0.5$, corresponding to an order of magnitude lower wind-feeding rate, similar to what a fit of the data to a spherically symmetric model would have yielded. The absorption in the latter model is shallower than the observed spectrum, even without any emission included, suggesting that such a rate is smaller than the actual wind-feeding rate of the star.

While this preliminary result does not allow us to place very strong constraints on the actual value of the wind-feeding rate, it does suggest that the observed UV spectrum of NGC 1624-2 is not inconsistent with the theoretical value computed by \citet{2012MNRAS.425.1278W}.




\section{Conclusions}\label{sec:concl}




NGC 1624-2 exhibits remarkable variations in the profiles of its wind-sensitive UV resonance lines. These changes are much larger than those seen in other magnetic massive stars. In particular, the variations in its C\textsc{iv} and Si\textsc{iv} offer interesting parallels with behaviours previously noted in other magnetic stars, but also show some distinct characteristics, in particular the redshifted absorption that can be seen in the low state spectrum as well as the double-peaked emission in Si\textsc{iv} in the high-state spectrum.


Compared to MHD simulations, 
the ADM model offers a computationally inexpensive alternative which is much better suited for large parameter studies. As such, deriving wind parameters using the ADM formalism becomes a tractable problem, as opposed to simply providing a qualitative explanation for the line morphology. Admittedly, there might be some important limitations associated with the assumptions made by the ADM model\footnote{One particular limitation of the ADM formalism, in the context of the faster-than-radial divergence which occurs at the pole and helps explain the strong line desaturation at high velocity, is that the resulting lower density along the magnetic pole is not taken into account dynamically, since a simple $\beta$-velocity law is used and no forces are being computed.}, but a more systematic study attempting to fit observational data with synthetic line profiles will help verify and refine these assumptions.


One difficulty in interpreting the variations seen in the spectra of various magnetic O stars comes from the uncertainties in the viewing angle of the magnetosphere ($\alpha$). While our current understanding is that the optical high state corresponds to the smallest viewing angle (closest to magnetic pole-on) and that the low state corresponds to the largest viewing angle (closest to magnetic equator-on), comparing these observations to models will require more precise determinations of these viewing angles, and therefore better magnetic characterizations of these stars. This is not an easy undertaking as they are all slow rotators, but dedicated observing runs to fill in the phase coverage will ultimately prove necessary to better reconcile data and theory.


Within the limited scope of this paper, we have used the ADM model to produce synthetic UV line profiles which we have then compared to the data and to a \textsc{cmfgen} model. The latter suggests that a low metallicity and microturbulent velocity are needed to reproduce the iron ``forests''. We have limited our modelling with the ADM to the absorption component of the line, which is suitable to reproduce line depth at high velocities, in an effort to quantitatively characterize the wind-feeding rate of NGC 1624-2. We find that using a non-spherically symmetric model, we can attribute the desaturation of the line at high velocity to the magnetospheric geometry without having to involve atypically low wind-feeding rates. This means that we find our high-state observation to be consistent with the theoretical wind-feeding rate inferred by \citet{2012MNRAS.425.1278W}. This constitutes a proof of concept as to the quantitative potential of the ADM formalism and its potential to yield a useful means of analysis of UV spectra for magnetic O stars, thus laying the foundation for upcoming efforts.

Many improvements to our current modelling tools are being developed, including the full treatment of both the ADM model (including the downflow and shock retreat components, and allowing for a finite Alfv\'{e}n radius) and radiative transfer (accounting for the ``emission'', or more accurately the light that is scattered into the line of sight). This will constitute the basis for an in-depth parameter study to investigate the full range of UV line profile behaviours (Erba et al., in prep.). These new developments will also help us validate and, as needed, fine-tune the ADM formalism for quantitative UV spectral analysis.

Further improvements could also take into account doublets, for a precise modelling of the C\textsc{iv} doublet for instance, as well as a varying line strength as a function of the ionization fraction throughout the wind.



\section*{Acknowledgments}

This paper is dedicated to the memory of our wonderful colleague and co-author, Dr. Nolan Walborn, who recently passed away. Much of this study would not have been possible had it not been for his pioneering work in spectral classification and on the Of?p class of magnetic massive stars, as well as his constant efforts to obtain UV spectroscopy of these stars. 

The authors would like to thank the reviewer (Prof. I. Howarth) for his helpful comments.

This research is based on observations made with the NASA/ESA {\it Hubble Space Telescope}, which is operated by the Association of Universities for Research in Astronomy, Inc., under NASA contract NAGS 5-26555.
Support for {\hst} General Observer Program number GO-13734 was provided by NASA through a grant from the Space Telescope Science Institute.

Support for this work was provided by NASA through Chandra
Award G03-14017C issued
by the Chandra X-ray Observatory Center which is operated
by the Smithsonian Astrophysical Observatory for and behalf of
NASA under contract NAS8-03060.

ADU gratefully acknowledges the support of the \textit{Fonds qu\'{e}b\'{e}cois de la recherche sur la nature et les technologies} and of the Natural Science and Engineering Research Council (NSERC) of Canada, as well as support from program HST-GO-15066.001-A that was provided by NASA through a grant from the Space Telescope Science Institute, which is operated by the Association of Universities for Research in Astronomy, Inc., under NASA contract NAS 5-26555.

CE acknowledges graduate assistant salary support from the Bartol Research Institute in the Department of Physics, University of Delaware, as well as support from program HST-GO-13629.002-A that was provided by NASA through a grant from the Space Telescope Science Institute.

JMA acknowledges support from the Spanish Government Ministerio de Econom{\'\i}a, Industria y Competitividad (MINECO/FEDER) through grant AYA2016-75\,931-C2-2-P.

YN acknowledges support from the Fonds National de la Recherche Scientifique (Belgium), the Communaut\'e Fran\c{c}aise de Belgique, and the PRODEX \textit{XMM} contract.

AuD acknowledges support by NASA through Chandra Award numbers GO5-16005X and TM7-18001X issued by the Chandra X-ray Observatory Center which is operated by the Smithsonian Astrophysical Observatory for and on behalf of NASA under contract NAS8-03060.

GAW acknowledges Discovery Grant support from NSERC.


\bibliographystyle{mn2e_fix2}
\bibliography{database}


\appendix


\section{List of observations}

The table below provides a list of all the UV spectroscopic observations of non-magnetic and magnetic OB stars shown in Fig.~\ref{fig:compNormal} and Fig.~\ref{fig:compMag}, respectively.


\begin{table*}
\caption[List of observations]{Full list of IUE and HST observations of the comparison non-magnetic and magnetic O stars used to produce, respectively, Fig.~\ref{fig:compNormal} and Fig.~\ref{fig:compMag}.}\label{tab:app}
\centering
\begin{tabular}{|c|c|c|c|c|c|c|c|}
	\hline
Star & Data ID & Telescope & Instrument & Exposure & Observation & Phase & Program ID and PI\\
     &         &           &            & time (s) & start time (UT) & & \\
	\hline
HD 93146 & SWP11136 & IUE & SWP & 3600 & 1981-01-24 01:43:07 & N/A & OD40B (J.E. Hesser) \\
\hline
HD 36861 & SWP46257 & IUE & SWP & 20 & 1992-11-13 06:25:36 & N/A & SWOJN (J.S. Nichols) \\
\hline
HD 108   & SWP08352 & IUE & SWP & 3300 & 1980-03-24 23:06:54 & $\sim 0.0$? & NSCAD (A.K. Dupree)\\
         & OBIL05010 & HST & STIS & 700 & 2010-09-09 16:15:11 & $\sim 0.5$? & 12179 (J.-C. Bouret)\\
\hline
HD 191612 & OBIL01010 & HST & STIS & 1500 & 2010-08-23 01:53:54 & 0.751 & 12179 (J.-C. Bouret)\\
          & OBIL02010 & HST & STIS & 1500 & 2011-01-04 06:49:30 & 0.000 & 12179 (J.-C. Bouret)\\
          & OBIL03010 & HST & STIS & 1500 & 2011-05-18 06:38:32 & 0.250 & 12179 (J.-C. Bouret)\\
          & OBIL04010 & HST & STIS & 1500 & 2011-09-30 12:02:56 & 0.501 & 12179 (J.-C. Bouret)\\
\hline
CPD -28 2561 & OCI7A1010 & HST & STIS & 2110 & 2014-04-27 20:42:35 & 0.01 & 13629 (Y. Naz\'{e})\\
             & OCI7A3010 & HST & STIS & 2000 & 2014-05-15 07:58:15 & 0.25 & 13629 (Y. Naz\'{e})\\
             & OCI7A2010 & HST & STIS & 1850 & 2014-06-01 20:11:45 & 0.49 & 13629 (Y. Naz\'{e})\\
\hline
HD 37022 & SWP13737 & IUE & SWP &   60 & 1981-04-17 01:31:52 & 0.756 & HSDRP (R.J. Panek)\\
         & SWP14665 & IUE & SWP &   90 & 1981-08-05 19:12:21 & 0.936 & LB304 (L. Bianchi)\\
         & SWP54001 & IUE & SWP &  120 & 1995-03-01 03:49:56 & 0.256 & RA110 (O. Stahl)\\
         & SWP54040 & IUE & SWP &  100 & 1995-03-05 03:36:57 & 0.513 & RA110 (O. Stahl)\\
         & SWP54112 & IUE & SWP &  100 & 1995-03-13 04:06:57 & 0.033 & RA110 (O. Stahl)\\
\hline
HD 57682 & SWP03576 & IUE & SWP &  356 & 1978-12-12 06:29:00 & 0.671 & RSLWK (L.W. Kamp)\\
         & SWP21587 & IUE & SWP &  180 & 1983-11-20 07:51:44 & 0.050 & MLFCG (C.D. Garmany)\\
\hline
HD 148937 & SWP02893 & IUE & SWP & 3600 & 1978-10-09 05:56:00 & 0.462 & OSPSC (P.S. Conti)\\
          & O6F301010 & HST & STIS & 883 & 2001-09-16 20:39:44 & 0.959 & 9243 (A. Boggess)\\
\hline
HD 54879  & OD3J01010 & HST & STIS & 383 & 2016-04-30 04:01:32	 & ?? & 14480 (W.-R. Hamann) \\
\hline
\end{tabular}
\end{table*}

\label{lastpage}
\end{document}